\documentclass[a4paper]{article}

\usepackage{icrc2013}
\usepackage[english]{babel}

\title{FACT - Long-term stability and observations during strong Moon light}

\shorttitle{FACT - Long-term stability and observations during strong Moon light}

\newcommand{\ethz}{$^1$}
\newcommand{\tudo}{$^2$}
\newcommand{\uniw}{$^3$}
\newcommand{\epfl}{$^4$}
\newcommand{\unige}{$^5$}

\authors{
M.~L.~Knoetig\ethz,
A.~Biland\ethz,
T.~Bretz\ethz,
J.~Bu\ss\tudo,
D.~Dorner\uniw,
S.~Einecke\tudo,
D.~Eisenacher\uniw,
D.~Hildebrand\ethz,
T.~Kr\"ahenb\"uhl\ethz,
W.~Lustermann\ethz,
K.~Mannheim\uniw,
K.~Meier\uniw,
D.~Neise\tudo,
\mbox{A.-K.}~Overkemping\tudo,
A.~Paravac\uniw,
F.~Pauss\ethz,
W.~Rhode\tudo,
M.~Ribordy\epfl,
T.~Steinbring\uniw,
F.~Temme\tudo,
J.~Thaele\tudo,
P.~Vogler\ethz,
R.~Walter\unige,
Q.~Weitzel\ethz,
M.~Z\"anglein\uniw $\;\;$
(FACT Collaboration)
}
\afiliations{
\ethz ETH Zurich, Switzerland $\;\;-\;\;$
  Institute for Particle Physics, Schafmattstr.~20, 8093 Zurich\\
\tudo Technische Universit\"at Dortmund, Germany $\;\;-\;\;$
  Experimental Physics 5, Otto-Hahn-Str.~4, 44221 Dortmund\\
\uniw Universit\"at W\"urzburg, Germany $\;\;-\;\;$
  Institute for Theoretical Physics and Astrophysics,
  Emil-Fischer-Str.~31, 97074 W\"urzburg,\\
\epfl EPF Lausanne, Switzerland  $\;\;-\;\;$
  Laboratory for High Energy Physics, 1015 Lausanne\\
\unige University of Geneva, Switzerland $\;\;-\;\;$
  ISDC Data Center for Astrophysics, Chemin d'Ecogia~16, 1290 Versoix \\
}

\email{mknoetig@phys.ethz.ch}

\abstract{ The First G-APD Cherenkov Telescope (FACT) is the first Cherenkov 
telescope equipped with a camera made of silicon photon detectors (G-APD aka.
SiPM). Since October 2011, it is regularly taking data on the Canary Island 
of La Palma. G-APDs are ideal detectors for Cherenkov telescopes as they are 
robust and stable. Furthermore, the insensitivity of 
G-APDs towards strong ambient light allows to conduct observations during 
bright Moon and twilight. This gain in observation time is essential for the 
long-term monitoring of bright TeV blazars.
During the commissioning phase, hundreds of hours of data (including data from the the Crab Nebula)
were taken in order to understand the performance and sensitivity of the 
instrument. The data cover a wide range of observation conditions including 
different weather conditions, different zenith angles and different light 
conditions (ranging from dark night to direct full Moon). We use a new parmetrisation 
of the Moon light background to enhance our scheduling and to monitor the atmosphere.
With the data from 1.5 
years, the long-term stability and the performance of the camera during Moon 
light is studied and compared to that achieved with photomultiplier tubes so 
far.}


\begin{document}
\maketitle

\section{Introduction}

\begin{figure}[th!]
  \centering
  \includegraphics[width=0.35\textwidth]{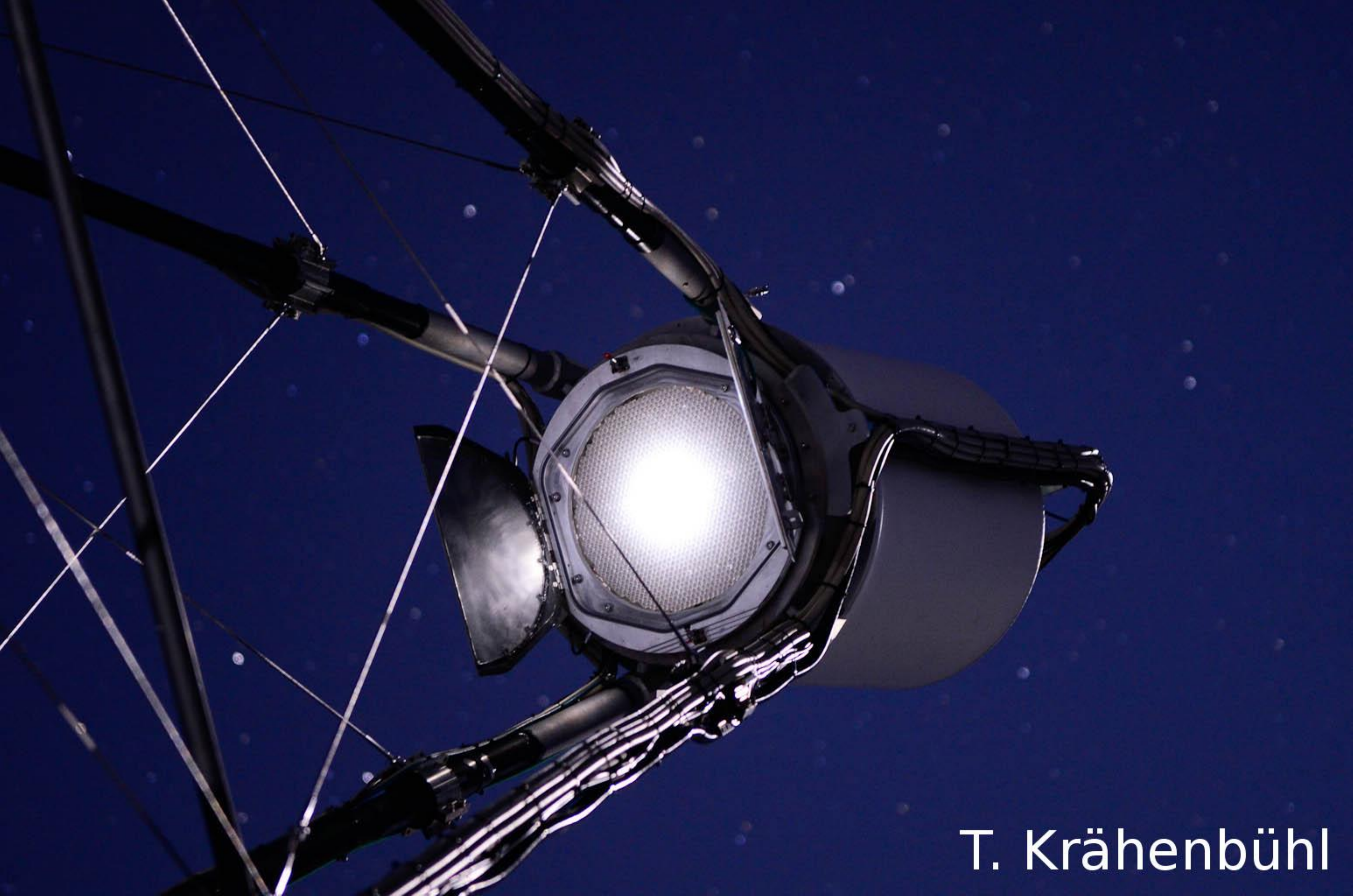}
  \includegraphics[width=0.30\textwidth]{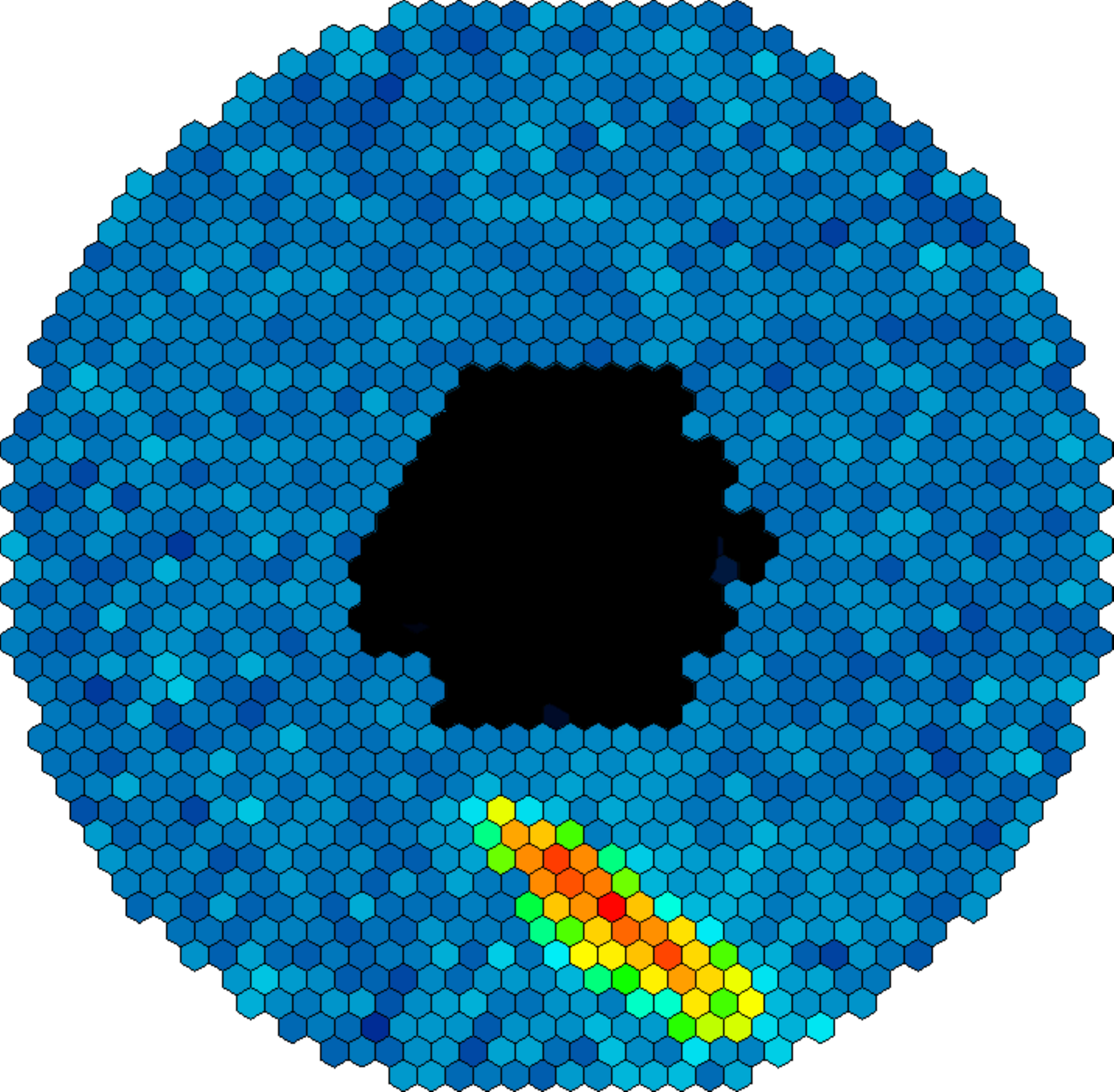}
  \caption{Top: The FACT telescope tracking the full Moon. Bottom: An event while 
  tracking the full Moon on the 23. June 2013. For this measurement, the power
  for the central patches was disabled.}
  \label{fig:Moon}
\end{figure}

The First G-APD Cherenkov Telescope (FACT) on the Canary Island of La Palma is the first Cherenkov 
telescope using novel silicon light detectors called Geiger-mode avalanche 
photo diodes (G-APD) or silicon photomultipliers (SiPM) instead of 
photomultipliers. These new detectors are compact, robust and can be used - 
unlike their predecessor - during strong ambient light. Since October 
2011, the telescope is operated at the Observatorio del Roque de los Muchachos 
(ORM, 2200m a.s.l.)\cite{bib:anderhub} and hundreds of hours of observations 
were conducted in order to understand the system. The background light 
conditions during datataking ranged from dark night to full Moon, and even the 
full Moon was once tracked in order to study response of the G-APD camera 
during the most extreme light conditions (Fig. \ref{fig:Moon}).

FACT is, since summer 2012, also the first remotely operated Cherenkov and will 
eventually be the first fully robotic telescope of it's kind
\cite{bib:adrian}. This is largely possible due to the stability and robustness 
of the novel light detectors. All of these properties benefit the main goal of 
FACT, which is to continuously monitor TeV Blazars over several years, in 
order to understand their variability and flaring behaviour. In the 
commissioning phase, FACT has already started to collect
large datasets from selected TeV Blazars\cite{bib:daniela}.

\section{Strong background light - technical considerations}
Today’s Cherenkov telescopes are equipped with light concentrators, 
which prevents most of the ambient light from entering the 
focal plane detector. The detected night sky background in the field of view 
of the camera is, for a dark night in the visible band, typically 
\mbox{$1.7\cdot10^{12}$ph m$^{-2}$ s$^{-1}$ sr$^{-1}$}\cite{bib:magic}.

When the Moon rises above the horizon, the situation changes.
The amount of scattered Moon light in the field of view depends on the 
angular distance of the observer to the Moon, the zenith positions of the 
source and the Moon, and with the Moon phase. Furthermore the atmospheric 
conditions can rapidly change the amount of scattered Moon light 
\cite{bib:krisciunas,bib:doro}. The amount of background light can go up by about a 
factor 1000, when looking close to the full Moon. These considerations equally 
apply to the few minutes of twilight every night. 

When observing with background light, one has to consider the possible damage 
to the light detectors. For photomultipliers this means the degradation of the
last dynode\cite{bib:magic}. 
In the past, observations of Cherenkov telescopes during Moon time were using UV 
filters\cite{bib:whipple}, solar blind photomultipliers, or a lowered 
photomultiplier high-voltage\cite{bib:hegra}, in order to reduce the amount of 
current coming from the scattered Moon light. Today’s Cherenkov telescopes use 
UV filters\cite{bib:veritas} and experiment with lower high-voltage, or 
standard voltage settings and higher trigger thresholds\cite{bib:magic} for 
Moon light and twilight observations. Around full Moon there is a gap of few 
days, where observations with current telescopes is not feasible\cite{bib:magic}. 

These above mentioned approaches are either time consuming, or expensive. By 
using G-APD as light detectors the constraint vanishes, as the 
G-APDs can be operated without serious degradation even during the brightest 
nights. Only the trigger condition has to be adjusted, depending on the 
changing light conditions, in order to keep the accidental trigger rate low. 

On the other hand, G-APDs can saturate when illuminated, as their number of 
cells is limited. Therefore one has to keep in mind their dynamic range, but
also the dynamic range of the digitisation and the trigger system. Lastly, 
when observing under strong background light conditions, the focal plane can 
heat up significantly because of the high DC currents. This is a challenge
for the bias feedback system, which has to keep the temperature dependent
overvoltage constant.

\section{G-APD stability}
A feature of G-APDs is that the breakdown voltage changes with the temperature.
For this reason the FACT bias feedback initially changed the bias voltage according
to the temperature coefficient specifications by the manufacturer, in order to achieve a constant gain. 
But also the DC current through the G-APDs has to be corrected for. From the
20. April 2012, the improved FACT bias feedback was comissioned, using also the 
DC current. 

In order to prove the long term G-APD feedback stability, a night by night 
extraction of the G-APD dark count spectrum was done.
This is possible, because G-APDs exhibit thermal induced breakdowns. In 
combination with the probability for crosstalk, this leads to a measurable 
signal spectrum, even when the FACT lid is closed \cite{bib:anderhub}, from which the 
individual pixel gain can be measured. The FACT dark 
rate is no problem during regular datataking, as it is of the order of MHz 
per G-APD, which a factor of ten less than the background rate from a dark night sky. 
 
\begin{figure}[t]
  \centering
  \includegraphics[width=0.49\textwidth]{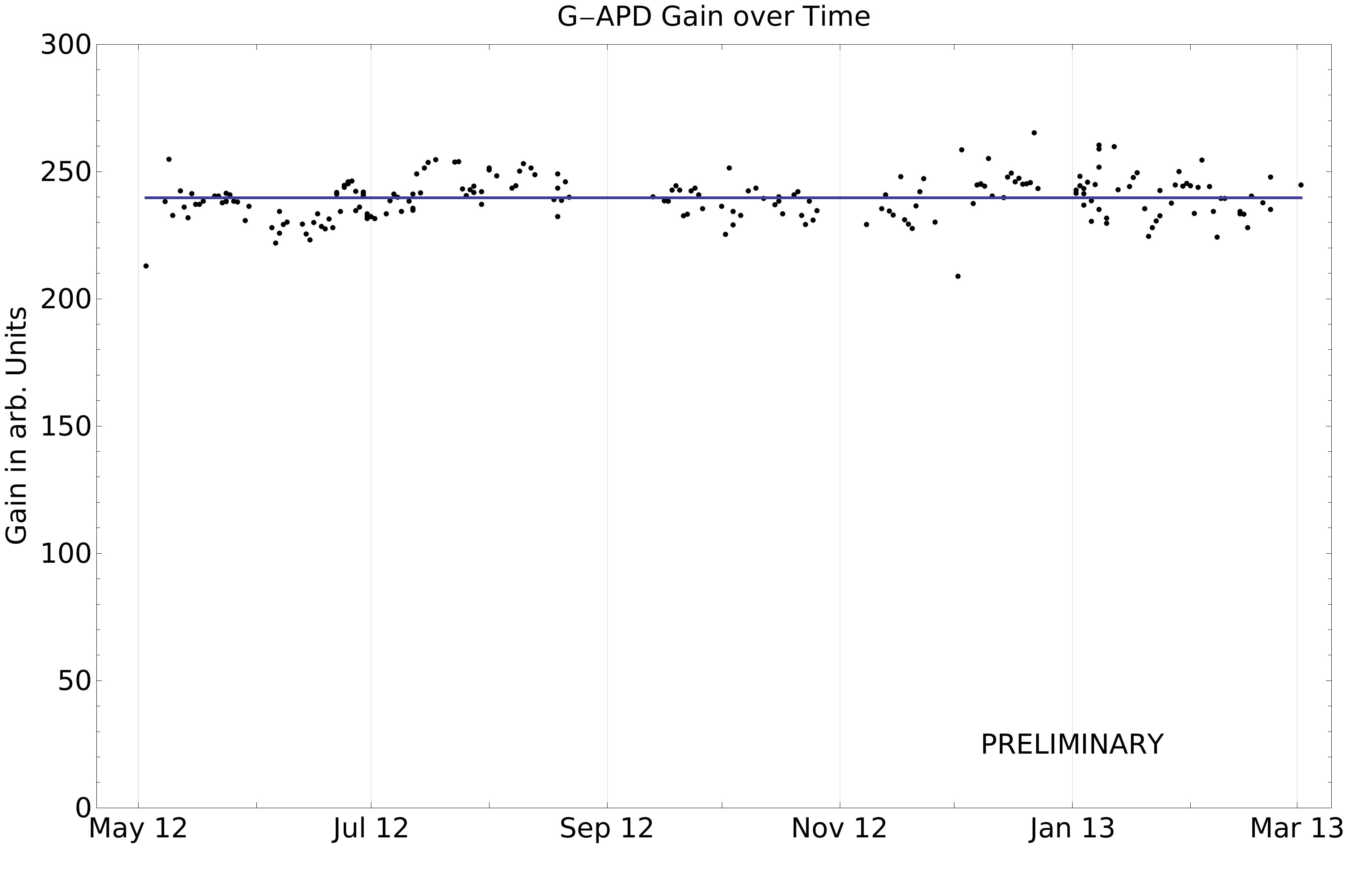}
  \caption{Mean G-APD gain over time. In April the improved feedback system was comissioned.}
  \label{fig:eff}
 \end{figure}

The results can be seen in Fig.\ref{fig:eff}.  Our data from May 2012 until March 2013 indicate
that the mean camera gain is stable with a standard deviation of $\sigma_t \simeq 3.4\%$. From a similar analysis
we deduce that our gain homogenity in the camera is $\sigma_{pix} \simeq 4\%$. 

As shown in \cite{bib:doro}, the rate of triggered Cherenkov
flashes for high enough threshold is independent of 
ambient light condition if the bias voltage is correctly
adjusted.

The method to verify the calibration relies on dark counts and crosstalk only and is
simpler, cheaper and more reliable than an external calibrated light source, as needed 
for photomultiplier cameras. These data show that it is possible to reliably 
calibrate a G-APD camera over many months using nothing but the ``off the shelf''
G-APD datasheet.

\section{Prediction of the brightness of Moon light}
The Moon illuminates the earth 
with reflected sun light. Krisciunas and Schaeffer proposed\cite{bib:krisciunas} 
a model for the calculation of the amount of photons detected, depending on 
the angular separation of the observer to the Moon, the zenith distance of the source and the 
Moon, and on the Moon phase. 

\begin{figure}[!t]
  \centering
  \includegraphics[width=0.49\textwidth]{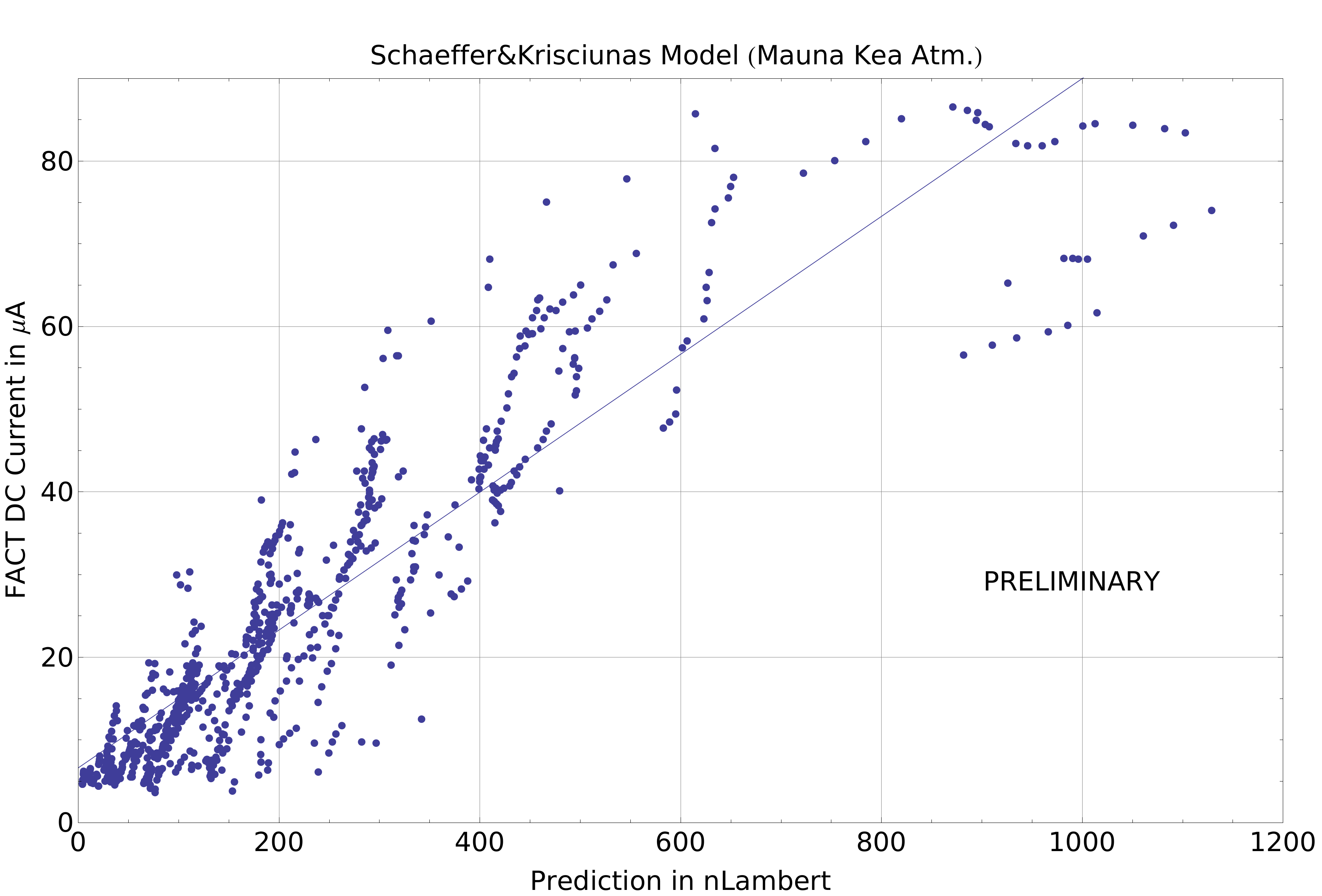}
  \includegraphics[width=0.49\textwidth]{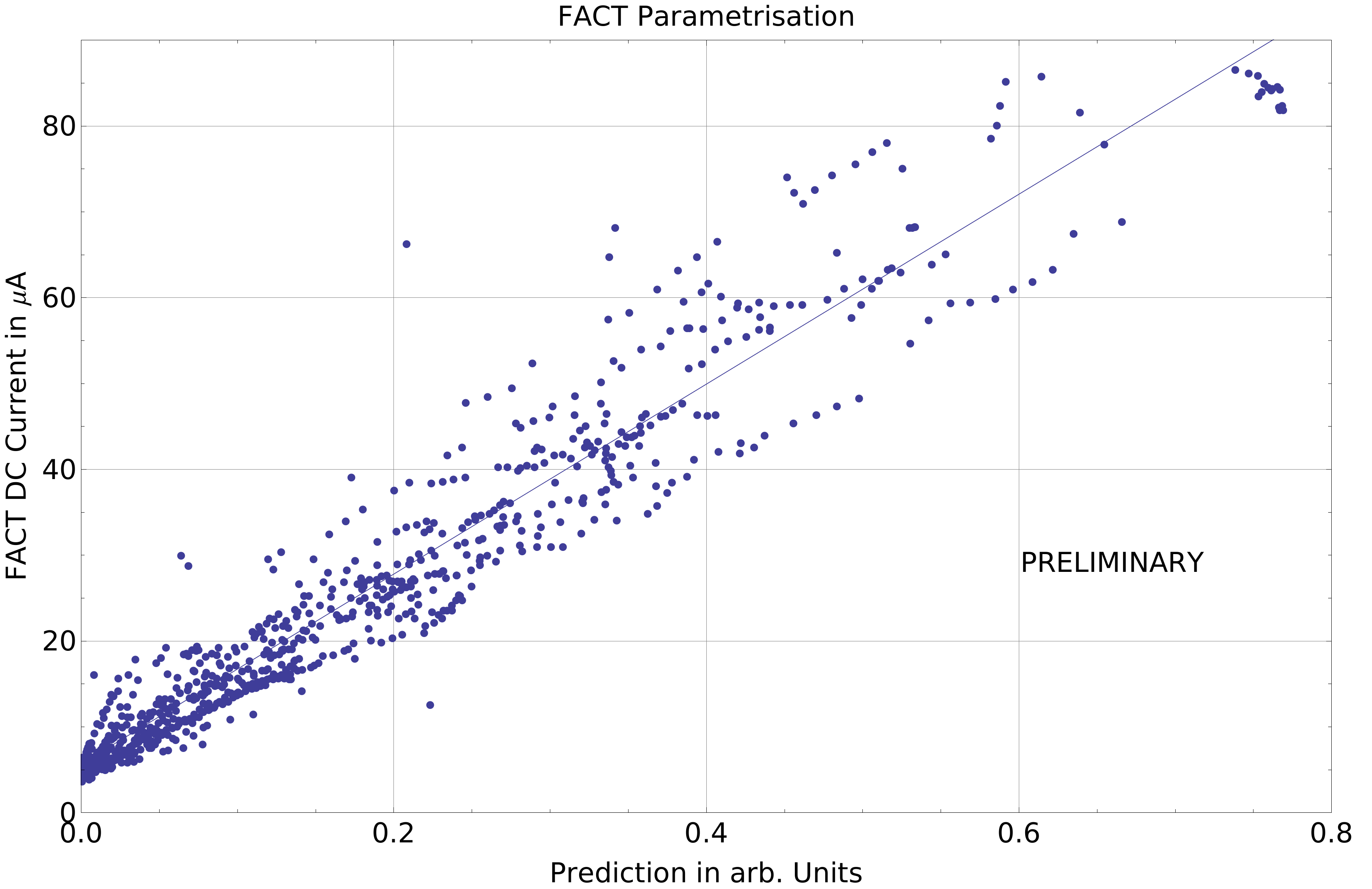}
  \caption{Moon light prediction models. Top: the model by Krisciunas and Schaeffer\cite{bib:krisciunas}. Bottom: simple empirical model. Only good weather datapoints included.}
  \label{fig:models}
 \end{figure}

Unfortunately, comparing the measured currents to this model shows large 
spread (Fig. \ref{fig:models} top). It is therefore 
remarkable that a simple empirical formula could be found to explain the DC 
currents during observations better(Fig. \ref{fig:models} bottom), where we define the Light Condition LC, 
dependent on the zenith distance of the 
Moon $Z_{Moon}<90^{\circ}$, and the Moon phase expressed as the illuminated 
fraction $A$ as 

\begin{eqnarray}
LC\left(Z_{Moon},A\right) & = & cos(Z_{Moon})\cdot A^{2.5}
\end{eqnarray}

Which is proportional to the DC current in the camera. 
The two linear parameters of the model can be extracted from a linear regression 
to the data. Including the zenith angle of the observer or the angular distance
to the Moon into the model does 
not seem to influence the results much for observations $>15^{\circ}$ away from the moon.
This is in agreement with the findings of others \cite{bib:krisciunas,bib:hegra,bib:magic}
It turns out that for small angular distances 
between Moon and observer, Mie-scattering dominates, and that for angular 
distances bigger than $\sim30^{\circ}$ a relatively constant background 
illumination can be expected from Rayleigh scattering\cite{bib:krisciunas,bib:hegra}.

The prediction of the brightness of Moon light has become an important 
ingredient, in order to successfully schedule observations.
With this model and its accuracy it is now for the first time possible to 
detect significant deviations from the predicted current, especially when a 
cloud moves through the field of view.

\section{Effect of background light on the telescope performance}
The much higher ambient light drastically increases
the rate of accidental triggers. Experience
with FACT show that for ~90\% of moon, the trigger
threshold has to be increased by about a factor of
three \cite{bib:thomas}. How much
this affects the analysis threshold is currently
under investigation, using a dataset of 363h of
data taken from the Crab nebula under varying
conditions during winter 2012/13.

\section{Conclusions}
With a G-APD camera, the calibration light source becomes obsolete, as 
the G-APDs are reliably pre-calibrated. By measuring the dark spectrum they 
can even be used to verify their own calibration. The data show a stable 
calibration since the improvement of the feedback in April 2012

The FACT telescopes using G-APDs as light detectors can, as the first Cherenkov 
telescope, observe with strong background light up to the full Moon without 
any problems or modifications to the datataking routine. The FACT G-APDs have 
been proven to work under direct full Moon illumination, which is many orders 
of magnitudes brighter than the dark night sky.

The sensitivity of the telescope is reduced by the increased trigger threshold and 
therefore energy threshold, which is needed to counteract accidental triggers.

In the future, we are planning to apply correction factors, obtained from 
the Crab Nebula, in order to determine the flux of $\gamma$-rays for any source zenith distance
and any light condition, in addition to Monte Carlo verifications.

\vspace*{0.5cm}
\footnotesize{{\bf Acknowledgment:}{ The important contributions from ETH 
Zurich grants ETH-10.08-2 and ETH-27.12-1 as well as the funding by the German 
BMBF (Verbundforschung Astro- und Astroteilchenphysik) are gratefully 
acknowledged. We are thankful for the very valuable contributions from E. 
Lorenz, D. Renker and G. Viertel during the early phase of the project. We 
thank the Instituto de Astrofisica de Canarias allowing us to operate the 
telescope at the Observatorio Roque de los Muchachos in La Palma, the 
Max-Planck-Institut f\"ur Physik for providing us with the mount of the former 
HEGRA CT\,3 telescope, and the MAGIC collaboration for their support. We also 
thank the group of Marinella Tose from the College of Engineering and 
Technology at Western Mindanao State University, Philippines, for providing us 
with the scheduling web-interface.}}

\begin{thebibliography}{}

\bibitem{bib:anderhub} H.~Anderhub {\it et al.} [FACT Collaboration], JINST (2013) {\bf 8}, P06008
\bibitem{bib:adrian} A.~Biland {\it et al.}  [FACT Collaboration], these proceedings, ID: 708
\bibitem{bib:daniela} D.~Dorner {\it et al.}  [FACT Collaboration], these proceedings, ID: 686
\bibitem{bib:magic} J.~Albert {\it et al.}  [MAGIC Collaboration], astro-ph/0702475 
\bibitem{bib:krisciunas} Krisciunas, K., \& Schaefer, B.~E.\, PASP, (1991) {\bf 103}, 1033 
\bibitem{bib:doro} D.~Hildebrand {\it et al.}  [FACT Collaboration], these proceedings, ID: 709
\bibitem{bib:whipple} {Chantell}, M. {\it et al.}, 24nd ICRC, Rome, (1995) {\bf 2}, 544
\bibitem{bib:hegra}  D.~Kranich {\it et al.}  [HEGRA Collaboration], Astropart. Phys., (1999)  {\bf 12}, 65
\bibitem{bib:veritas} D.~Staszak for the VERITAS Collaboration, 4th RICAP, Rome, 2013
\bibitem{bib:thomas} T.~Bretz {\it et al.}  [FACT Collaboration], these proceedings, ID:720

\end{thebibliography}
\end{document}